# Preventing Phishing Attacks using One Time Password and User Machine Identification

Ahmad Alamgir Khan
Senior IT Consultant,
Muscat Securities Market
Muscat, Oman

## ABSTRACT
Phishing is a type of attack in which cyber criminals tricks the victims to steal their personal and financial data. It has become an organized criminal activity. Spoofed emails claiming to be from legitimate source are crafted in a way to lead victims to reveal their personal, financial data by misdirecting them to the counterfeit website.

This research paper presents a novel approach to combat the Phishing attacks. An approach is proposed where user will retrieve the one time password by SMS or by alternate email address. After receiving the one time password the web server will create an encrypted token for the user's computer/device for authentication. The encrypted token will be used for identification, any time user wishes to access the website he/she must request the new password. The one time password as name implies will expire after single use. The one time password and encrypted token is a smart way to tackle this problem.

## General Terms
Phishing prevention, Encryption, SMS, One Time Password et. al.

## Keywords
Phishing attacks and prevention, Anti phishing, SMS, One Time Password, OTP, Authentication, X509Certificate2, Encryption, Client Identity and APPT.

## 1. INTRODUCTION
Phishing mostly uses spoofed e-mail messages that seem to come from legitimate source. Trojans, malware and other malicious software are also used for phishing attacks. Since, SMTP protocol does not validate or authenticate the sender of the email; anyone can claim to be a valid sender such as from banks, credit card companies and other agencies. Cyber criminals make sure that the message appears as trustworthy as possible that includes formats similar to the legitimate message. By using such social engineering techniques they are able to allure the victims to respond to those messages.

According to the Russell Kay [3], up to 20% of unsuspecting recipients may respond to them, resulting in financial losses, identity theft and other fraudulent activity against them. Financial institutions are at risk for large numbers of fraudulent transactions using the stolen information [4].

The absence of inherent security features in the current web technologies have made it possible for attackers to exploit the flaws and take the significant advantage. There are several techniques to combat such attacks yet none have proven to be 100% full proof due to the fact that the criminals can easily find one or other way to deceive the victims.

According to Dr. Dobb's [1] more than 500 million phishing e-mails appear in user inboxes every day.

According to the APWG [2], the word "phishing" appeared around 1995, when Internet scammers were using email lures to "fish" for passwords and financial information from the sea of Internet users; "ph" is a common hacker replacement of "f", which comes from the original form of hacking, "phreaking" on telephone switches during 1960s.

The impact of phishing on the global economy has been quite significant: RSA estimates that worldwide losses from phishing attacks cost more than $1.5 billion in 2012, and had the potential to reach over $2 billion if the average uptime of phishing attacks had remained the same as 2011 [5].

In this paper, the Anti-Phishing prevention technique (APPT) using one time password (OTP) and encrypted token aims to protect users against spoofed web site-based phishing attacks. It has been designed using .NET 3.5 and SQL Server 2005.

The paper is structured as follows: Section 2 provides a brief overview of the different types of phishing attacks. Section 3 describes the different phishing techniques. Section 4 presents an example that shows how the APPT works, and provides details about its implementation and shows how it mitigates a typical phishing attempt. Section 5 presents future work and concludes the article.

## 2. TYPES OF PHISHING ATTACKS
Phishing attacks are not limited to spoofed emails only; it includes search engines, man-in-middle, malware, Trojans, instant messaging, social networking sites and etc. Below are some major categories of phishing.

### 2.1 Clone Phishing
It is a type of an attack where a legitimate previously delivered email containing an attachment o-r link has had its content and recipient address (es) taken and used to create a cloned email.

The attachment or link within the email is replaced with a malicious version and then sent from an email address spoofed to appear to come from the original sender [6]. It may claim to be a re-send of the original or an updated version to the original. This technique could be used to pivot (indirectly) from a previously infected machine and gain a foothold on another machine, by exploiting the social trust associated with the inferred connection due to both parties receiving the original email [7].

### 2.2 Spear Phishing
It is a technique where specific victim is targeted. The information about the victim is known prior to the attack and the email is sent from the source known by the victim. Due to the nature of the trust on receiving email, this kind of attack has high probability to be successful. An example would be receiving an email from friend, colleague or financial institutions which prompt victim to provide the credentials.

## 3. PHISHING TECHNIQUES

Many techniques are developed to conduct phishing attacks. Malicious person with novice computer skills can use tools which are available freely on the internet to conduct a devastating phishing attack and make them less susceptible.

Web Spoofing is a method in which forged website looks similar to the legitimate one so that users can enter their confidential information. Email spoofing may occur in different forms, but all have a similar result: a user receives email that appears to have originated from one source when it actually was sent from another source. Email spoofing is often an attempt to trick the user into making a damaging statement or releasing sensitive information (such as passwords) [8].

### 3.1 Web Spoofing

Web Spoofing is a security attack that allows an adversary to observe and modify all web pages sent to the victim's machine, and observe all information entered into forms by the victim. Web Spoofing works on both of the major browsers and is not prevented by "secure" connections. The attacker can observe and modify all web pages and form submissions, even when the browser's "secure connection" indicator is lit. The user sees no indication that anything is wrong [9].

Once this information is collected, the attacker can use it to buy things with the victims' credit cards, access their bank accounts, and establish false identities. Website spoofing is a growing phenomenon, and puts consumers at considerable risk for identity theft and credit card fraud [10].

The attack is initiated when the victim visits a malicious Web page, or receives a malicious email message (if the victim uses an HTML-enabled email reader). [9]

### 3.2 E-mail spoofing

Email spoofing is email activity in which the sender address and other parts of the email header are altered to appear as though the email originated from a different source. Because core SMTP doesn't provide any authentication, it is easy to impersonate and forge emails [11]. Distributors of spam often use spoofing in an attempt to get recipients to open, and possibly even respond to, their solicitations. Spoofing can be used legitimately. Classic examples of senders who might prefer to disguise the source of the e-mail include a sender reporting mistreatment by a spouse to a welfare agency or a "whistle-blower" who fears retaliation [12].

The focus of this paper is to use the above mentioned information to prevent the attacks using APPT.

## 4. ANTI-PHISHING TECHNIQUE (APPT)

Many solutions have been developed to combat the phishing attacks. It includes both technical and non-technical target areas. Yet, the tendency of phishing is expanding and the numbers of new techniques are implemented to cause more damage, it has become a serious cybercriminal activity.

Anti-Phishing Prevention Technique namely APPT is based on the concept of preventing phishing attacks by using combination of one time random password and encrypted token for user machine identification. First step is to retrieve the password by SMS or by alternate emails, during that process encrypted token is created which have user specific data and is stored in the user machine. Second step is to access the required website with the password and valid token which are required for successful authentication. This Section presents the implementation and how it can help in mitigating a typical phishing attempt.

Figure 1. Getting one time password and identifying machine

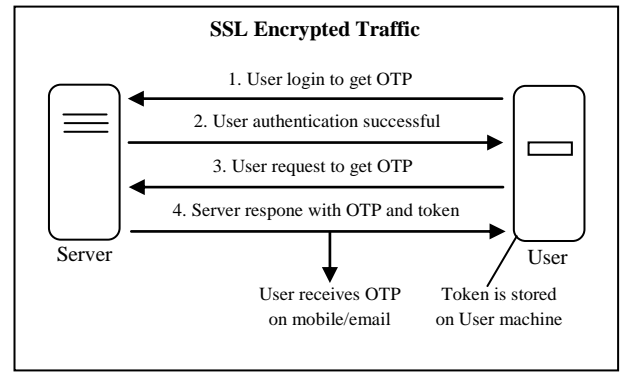

### 4.1 Getting One Time Password

1. Users will go to the one time password retrieval site to receive the random password.

https://www.mysite.com/getpassword.aspx

| APPT - Get One Time Password | |
|---|---|
| User Name: | |
| Password: | |
| Login | |

User Login Table for One Time Password

| User Name | Password | Mobile No. | Email |
|---|---|---|---|
| aqwert | P2323!23 | 9689563581 | aqwr@yml.co |
| twerffy | 6534g&* | 9689012561 | twer@yml.co |
| ttyuuii | P#$2334 | 9669525452 | rtyu@yml.co |
| yuhfry | Ad#$%8 | 9689123565 | yuhf@yml.co |
| ikioljui | rT%^$11 | 9678595231 | ikio@yml.co |

2. After authentication, users can retrieve the one time password either by SMS or Email. These records are already present in database along with the user credentials.

**APPT – Select option to get Password**

Send to SMS (more secure)

Send to Alternate Email Address

3. Clicking on the one of the button will create one time random password and will create an encrypted token on the user machine and in database on the Server. The password would be sent based on selection above.

Login Table to Main Website with Token Name

| User Name | Password | Status | Token Name |
|---|---|---|---|
| aqwert | 895$%6! | 1 | Random value |
| twerffy | Expired | 0 | Expired |
| ttyuuii | P#$2334 | 1 | Random value |
| yuhfry | Ad#$%8 | 1 | Random value |

- **Generate encrypted token, store in DB and send OTP**

*Encrypt Cookie Code*

*Getting the Certificate from the Store*

X509Store store = new X509Store(StoreName.Root, StoreLocation.LocalMachine);

store.Open(OpenFlags.ReadOnly);

X509Certificate2 Cert = store.Certificates.Find( X509FindType.FindByThumbprint, "08 15 b7 a7 26 d3 06 0a 4f 61 b9 eb f7 e4 0f 7a 7c 24 6f 0c",  false);

*Encryption Function*

public static string PKIEncrypt(string data, X509Certificate2 certificate)  {

  if (certificate == null)
     throw new ArgumentException("Certificate cannot be null.");

  if (certificate.HasPrivateKey == false)
     throw new CryptographicException("Certificate does not contain a private key.");

RSACryptoServiceProvider rsa = (RSACryptoServiceProvider) certificate.PublicKey.Key;

  byte[] plainbytes = Encoding.UTF8.GetBytes(data);
  byte[] cipherbytes = rsa.Encrypt(plainbytes, false);
  return Convert.ToBase64String(cipherbytes);  }

// when user gets the password, random cookie is generated
HttpCookie cookie = new HttpCookie("APPTSecureCookie");

// tokenName = "39369a0d-975e-44….." randomly generated
cookie["TokenName"] = tokenName;

cookie["HostUserName"] = username;  //current user
cookie["Email"] = userEmail; //user email from DB
cookie["HostIPAddress"] = Request.UserHostAddress;

cookie.Expires = DateTime.Now.AddMinutes(15);
cookie.HttpOnly = true;  // mitigates cross site attack
cookie.Value = Encryption.PKIEncrypt(cookie.Value, Cert);

Response.Cookies.Add(cookie);

// update to database, password=random password
DB.Save(username, password, status, tokenName);

// send SMS or Email
sendPassword(password);

*Explanation X509Certificate2^*

The X.509 structure originated in the International Organization for Standardization (ISO) working groups. This structure can be used to represent various types of information including identity, entitlement, and holder attributes (permissions, age, sex, location, affiliation, and so forth). Although the ISO specifications are most informative on the structure itself, the X509Certificate2 class is designed to model the usage scenarios defined in specifications issued by the Internet Engineering Task Force (IETF) Public Key Infrastructure, X.509 (PKIX) working group. [13]

### 4.1.1 Mitigation against the attack on one time password retrieval website

PKIEncrypt function is used to encrypt the token. The cookie is valid only for 15 minutes and contains user machine IP address and other details. The token stored on user machine is used along with the one time password to authenticate with the primary website.

***Forged Site***, *http://www.mysite.net/getpassword.aspx*

| **APPT - Get One Time Password** |
|---|
| User Name: [          ] |
| Password: [          ] |
| [ Login ]    [ **Forged website** ] |

If attacker creates a forged website for getting the one time password, he/she will not be able to cause any damage as the site is only used to retrieve the password on mobile or email which is only accessible to the valid user. In order to mitigate against the Cookie attack, the token (cookie) expires in 15 minutes and the cookie is transmitted over encrypted channel and is also encrypted with X509Certificate2 certificate.

However, if attacker is able to get the user credentials by forged website, it can provide an opportunity for an attacker to flood the victim with SMSs or with email messages and can create lot of nuisance. To prevent such flooding CAPTCHA is used.

A CAPTCHA is a program that protects websites against bots by generating and grading tests that humans can pass but current computer programs cannot. For example, humans can read distorted text, but current computer programs can't. [14]

### 4.2 Accessing the Sensitive Website

The figure 2 below mentioned demonstrates how User and Machine authentication is performed.

Figure 2. Authenticate User and machine token

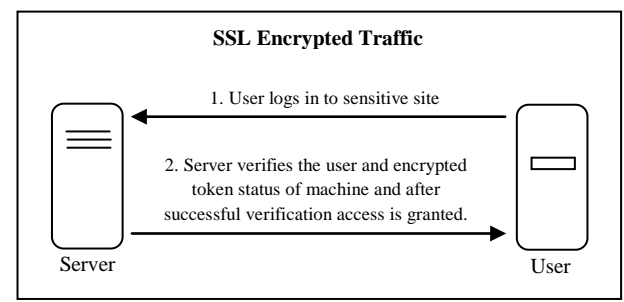

1. Users login to the Sensitive website.
https://www.mysite.com/shopping.aspx

**APPT – Sensitive Website Login**

User Name: aqwert

Password: *******

[Login]

2. User enters user name and password (one time password) and clicks on "Login"

3. The status of user is checked, the user name, password and token Name is retrieved from database.

Login Table to Main Website with Token Name

| User Name | Password | Status | Token Name |
|---|---|---|---|
| aqwert | 895$%6! | 1 | random value |
| twerffy | Expired | 0 | Expired |

4. Token (Cookie) is decrypted and is checked for validity.

- **Validate user & verify machine token status**

*// validate user from login table, returns token name*
tokenName = DB.ValidateUser (username, password);

*Getting the Certificate from the Store*

X509Store store = new X509Store(StoreName.Root, StoreLocation.LocalMachine);

store.Open(OpenFlags.ReadOnly);

X509Certificate2 Cert = store.Certificates.Find( X509FindType.FindByThumbprint, "08 15 b7 a7 26 d3 06 0a 4f 61 b9 eb f7 e4 0f 7a 7c 24 6f 0c",  false);

*Decrypt Token (Cookie) Function:*

public static string PKIDecrypt(string Base64EncryptedData, X509Certificate2 certificate)  {

if (certificate == null)
throw new ArgumentException("Certificate cannot be  null.");

if (certificate.HasPrivateKey == false)
   throw new CryptographicException("Certificate does
   not contain a private key.");

RSACryptoServiceProvider rsa = (RSACryptoServiceProvider)certificate.PrivateKey;

byte[] decryptedBytes =
       rsa.Decrypt(Convert.FromBase64String(
       Base64EncryptedData), false);

       return Encoding.UTF8.GetString(decryptedBytes);
}

*Get the Token (Cookie) values*

TokenValid = false;

*//sets the password expired after one time successful login*
DB.setPasswordExpired (username);

If (tokenName != null) {

HttpCookie cookie = context.Request.Cookies["APPTSecureCookie "];

cookie = PKIDecrypt (cookie); //decrypt

if (cookie != null && ! cookie.Expires && cookie.Secure ) {

   if (cookie["TokenName"] = tokenName &&
     cookie["HostUserName"] = username;  &&
     cookie["HostIPAddress"] = Request.UserHostAddress)
   {
      TokenValid = true;  }
   }
}

If (TokenValid )
   // TokenValid successful authentication
Else
   // Unsuccessful authentication

5. The above describes the process of getting the user authenticated from the database and verifies all the parameters of Token to verify the identity. Here cookie.Expires check for expiration and cookie.Secure makes sure that request is coming through secure channel and etc.

***Successful Login*** *- Valid user name, password, Token and other details.*

**APPT – Welcome User: aqwert**

Card Holder Name: 

Card Number: 

Expiry Date: 

Security Code:    [Pay]

## 4.2.1 Mitigation against the website attack

The login will only take place when the user will have valid token. The one time password is expired after successful login. The expiry time of token is set to 15 minutes, the token is decrypted by the certificate on the issuing server which makes it difficult to crack and alter in transit. As we can see in code, decrypted token is again checked for valid Host IP address and it is made sure that token is coming from the Secure connection. Also, during the creation of token, HttpOnly option is set to true which helps in mitigation of cross site scripting attacks.

In order to access the sensitive site it is necessary to have valid token and user credentials. An example if in case attacker gets the access to the one time password.

***Forged Site***, *http://www.mysite.net/shopping.aspx*

Figure 3. Attackers tries to Access the Sensitive Site

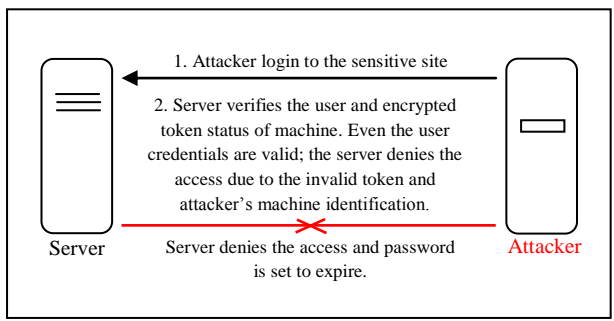

By creating a forged site the attacker mislead the victims to provide credentials (OTP). The user credentials are retrieved by the attacker in order to access the sensitive website. As in Figure 3, once attacker tries to access the website with the retrieved user credentials, the valid token (cookie) and other machine identification parameters are checked which are not valid and therefore access is denied. The user credentials are immediately set to expire in order to stop the replay attack. Due to expiry time of token and one time password it will be very difficult to launch a successful attack.

## 5. CONCLUSION AND FUTURE WORK

Phishing emails and web site attacks have provided a faceless opportunity for fraudsters to reach millions of potential victims, with little cost outlay, in the hope of victims supplying their personal and financially sensitive information. This information is then used to hijack accounts and duplicate the victim's identity through the fastest growing crime in the world, Identity Theft. It is apparent that fraudsters perpetrating phishing scams are becoming more technologically efficient, utilizing smarter deception methods to create and implement their phishing scams [15].

However, by using APPT it can be assured that attack like Phishing can be prevented to a large extent. However, future work has to be done to provide more secure encryption technology that will be difficult to break. Users also have to be made aware of the risks they face on internet, and they should responsibly use the internet for their benefit. APPT alone cannot help in mitigating all the types of Phishing attacks, as there are many ways attackers are using to commit the fraud. It is highly recommended that good antivirus/spyware/malware software be installed on user machines and are regularly updated. The operating system and other software security updates are very important too, they must be regularly updated. With due care of using internet user can reap the benefits of the internet technology. Users must understand that the attackers take benefits of the habitual usage of internet where users always never reply to any message with "no", they tend to like to click on "yes". Users should properly verify their actions and should know the consequence of them. It is important to be aware of dangers of such attacks and their devastating effects.